\documentclass[english,aps,manuscript]{revtex4-1}
\usepackage[T1]{fontenc}
\usepackage[latin9]{inputenc}
\usepackage[active]{srcltx}
\usepackage{amssymb}
\usepackage{esint}

\makeatletter
\@ifundefined{textcolor}{}
{%
 \definecolor{BLACK}{gray}{0}
 \definecolor{WHITE}{gray}{1}
 \definecolor{RED}{rgb}{1,0,0}
 \definecolor{GREEN}{rgb}{0,1,0}
 \definecolor{BLUE}{rgb}{0,0,1}
 \definecolor{CYAN}{cmyk}{1,0,0,0}
 \definecolor{MAGENTA}{cmyk}{0,1,0,0}
 \definecolor{YELLOW}{cmyk}{0,0,1,0}
 }

\makeatother

\usepackage{babel}
\begin{document}

\title{Area Spectrum of a Rotating Charged Black Hole Solution of Heterotic
String Theory}

\author{Alexis Larrañaga}

\address{National Astronomical Observatory. National University of Colombia.}
\begin{abstract}
The recent proposal of Maggiore that the periodicity of a black hole
may be the origin of area quantization law is analized in the context
of black holes in string theory. We use the period of motion of an
outgoing wave, which is shown to be related to the vibrational frequency
of the perturbed black hole, to quantize the horizon areas of a Sen
black hole. It is shown that the equally spaced area spectrum takes
the same form as the obtained by Zeng et. al. for Schwarzschild and
Kerr black holes and the spacing is the same as that obtained through
the quasinormal mode frequencies. In order to obtain this result,
we do not need to use the small angular momentum assumption which
is necessary in the quasinormal mode approach.


\end{abstract}
\maketitle

\section{Introduction}

Recently, investigation on quasinormal modes of black holes has shown
the possibility to find black holes by detecting their astrophysical
consequences \cite{Kokkotas1999} and also provided a way to check
the string theory \cite{Konoplya83}. Additionaly, it was shown that,
using Bohr's correspondence principle which states that transition
frequencies at large quantum numbers are equal to classical oscillation
frequencies, the quasinormal modes also provide a method to quantize
the horizon area of black holes. 

According to \cite{Hod1998}, the real part of quasinormal mode frequency
is responsible for the area spectrum of black holes. Based on the
quasinormal modes of Schwarzschild black hole in large $n$ limit
\cite{Nollert16,Leaver402}, and later using the adiabatic invariant
\cite{Kunstatter2003} they obtained the quantized horizon area 

\begin{equation}
\Delta A=32\pi M\delta M=4\ln3l_{p}^{2}
\end{equation}

where $M$ is the black hole mass and $l_{p}$ is Planck's lenght
(in units with $G=c=1$). This idea stems from an analogy with the
classical harmonic oscillator: the action integral of the form $A=\oint pdq$
for a quasiperiodic system is an adiabatic invariant in analytical
mechanics. Specifically, the Hamiltonian of the one-dimensional harmonic
oscillator is

\begin{equation}
H=\frac{p^{2}}{2m}+\frac{m\omega_{c}^{2}q^{2}}{2}
\end{equation}
and the corresponding adiabatic invariant is 
\begin{equation}
A=\frac{E}{\omega_{c}}.
\end{equation}
Therefore, in the black hole case the classical vibrational frequency
$\omega_{c}$ and the system energy $E$ are treated as the quasinormal
mode frequencies $\omega$ and the mass $M$ in the large $n$ limit,
giving 

\begin{equation}
A=\int\frac{dM}{\omega}.
\end{equation}

Maggiore \cite{Maggiore2008} proposed that this treatment should
be reexamined because the proper frequency of the equivalent harmonic
oscillator contains real $\omega_{R}$ and imaginary $\omega_{I}$
contributions. Since the imaginary contribution is dominant for the
highly excited quasinormal modes, one should use the imaginary part
of quasinormal mode frequencies to study the area spectrum of black
holes. This modification gives in the case of Schwarzschild black
hole, the quantized horizon area 
\begin{equation}
\Delta A=8\pi l_{p}^{2}
\end{equation}
 and the origin of this area quantization appears to be the periodicity
of the black hole in Euclidean time. The form of the horizon area
quantized in units of $l_{p}$ was proposed firstly by Bekenstein
\cite{Bekenstein} and he also found that the horizon area of a non-extremal
black hole is adiabatic invariant classically. Using the point particle
model presented in \cite{Christodoulou1970}, Bekenstein found that
the smallest possible increase in horizon area of a non-extremal black
hole is exactly $\Delta A=8\pi l_{p}^{2}$. 

Following the work of Maggiore, area spectrum of many black holes
has been investigated as for example rotating metrics in \cite{Vagenas0811,Medved25}
and charged black holes in \cite{Wei2010,Lopez-Ortega}. There are
also studies via quasinormal modes analysis in de Sitter spaces \cite{li676,Chen69},
non-Einstein gravity \cite{Kothawala104018,Majhi49,Banerjee279,Wei03,jiang2010}
and other background spacetimes \cite{Kwon27,Kwon27165011,Kwon282011,Daghigh26,Wei0903}.

Banerjee et. al. \cite{Banerjee2010}, found that area spectrum of
black holes can be obtained by computing the average squared energy
of the outgoing wave in the view of quantum tunneling and recently,
Zeng et. al. \cite{Zeng2012} employ the periodicity of outgoing wave
to obtain area spectrum of Schwarzschild and Kerr black holes. Their
idea is that for a perturbed black hole, the outgoing wave performs
periodic motion outside the horizon and the corresponding period is
related to the frequency of the outgoing wave. The gravity system
in Kruskal coordinates is periodic with respect to Euclidean time
and the motion of a particle in this periodic gravity system also
owns a period given by the inverse of the Hawking temperature. Therefore
they conclude that the frequency of the outgoing wave is given by
the inverse of the Hawking temperature.

In this paper, we will follow the treatment of Zeng et. al. to obtain
the area spectrum of the rotating charged black hole of heterotic
string theory reported by Sen \cite{Sen}. In order to obtain the
area spectrum, we get the concrete value of vibration frequency by
equaling the motion period of the outgoing wave to the period of the
gravity system obtained when considering the Euclidean time.

\section{The Sen black hole}

Sen \cite{Sen,Senotro} was able to find a charged, stationary, axially
symmetric solution of the field equations by using target space duality,
applied to the classical Kerr solution. The line element of this solution
can be written, in generalized Boyer-Linquist coordinates, as

\begin{eqnarray}
ds^{2} & = & -\left(1-\frac{2Mr}{\rho^{2}}\right)dt^{2}+\rho^{2}\left(\frac{dr^{2}}{\Delta}+d\theta^{2}\right)-\frac{4Mra\sin^{2}\theta}{\rho^{2}}dtd\varphi\nonumber \\
 &  & +\left(r\left(r+r_{\alpha}\right)+a^{2}+\frac{2Mra^{2}\sin^{2}\theta}{\rho^{2}}\right)\sin^{2}\theta d\varphi^{2},\label{eq:kerrsen}
\end{eqnarray}
where 

\begin{eqnarray}
\Delta & = & r\left(r+r_{\alpha}\right)-2Mr+a^{2}\\
\rho^{2} & = & r\left(r+r_{\alpha}\right)+a^{2}\cos^{2}\theta.
\end{eqnarray}

Here $M$ is the mass of the black hole, $a=\frac{J}{M}$ is the specific
angular momentum of the black hole and the electric charge is given
by

\begin{equation}
r_{\alpha}=\frac{Q^{2}}{M}.
\end{equation}
Note that in the particular case of a static black hole, i.e. $a=0$,
the metric (\ref{eq:kerrsen}) coincides with the GMGHS solution \cite{gmghs}
while in the particular case $r_{\alpha}=0$ it reconstructs the Kerr
solution.

The Kerr-Sen space has a spherical event horizon, which is the biggest
root of the equation $\Delta=0$ and is given by

\[
r_{H}=\frac{2M-r_{\alpha}+\sqrt{\left(2M-r_{\alpha}\right)^{2}-4a^{2}}}{2}
\]

or in terms of the black hole parameters $M,Q$ and $J$,

\begin{equation}
r_{H}=M-\frac{Q^{2}}{2M}+\sqrt{\left(M-\frac{Q^{2}}{2M}\right)^{2}-\frac{J^{2}}{M^{2}}}.\label{eq:horizon}
\end{equation}
The area of the event horizon is given by

\begin{equation}
A=\int\left(r_{H}^{2}+a^{2}\right)\sin\theta d\theta d\varphi=8\pi M\left(M-\frac{Q^{2}}{2M}+\sqrt{\left(M-\frac{Q^{2}}{2M}\right)^{2}-\frac{J^{2}}{M^{2}}}\right).\label{eq:area}
\end{equation}

Equation (\ref{eq:horizon}) tell us that the horizon disappears unless

\[
\left|J\right|\leq M^{2}-\frac{Q^{2}}{2},
\]
therefore, the extremal black hole, $\left|J\right|=M^{2}-\frac{Q^{2}}{2}$,
has $A=8\pi\left|J\right|$. The angular velocity at the horizon is
given by

\begin{equation}
\Omega=\frac{J}{2M^{2}}\frac{1}{M-\frac{Q^{2}}{2M}+\sqrt{\left(M-\frac{Q^{2}}{2M}\right)^{2}-\frac{J^{2}}{M^{2}}}}\label{eq:angularvelocity}
\end{equation}

its electrostatic potential at the horizon can be written as

\begin{equation}
V=\frac{Q}{2M}\label{eq:electricpotential}
\end{equation}
and the Hawking temperature is

\begin{equation}
T_{H}=\frac{\kappa_{H}\hbar}{2\pi}=\frac{\hbar\sqrt{\left(2M^{2}-Q^{2}\right)^{2}-4J^{2}}}{4\pi M\left(2M^{2}-Q^{2}+\sqrt{\left(2M^{2}-Q^{2}\right)^{2}-4J^{2}}\right)}.\label{eq:HawkingTemperature}
\end{equation}

For the Sen black hole, there is an ergosphere between the outer horizon
and the infinite redshift surface. To avoid the dragging effect, one
should perform the dragging coordinate transformation\cite{Chen665},
\begin{equation}
\phi=\varphi-\Omega t,
\end{equation}
where the dragging angular velocity is given by (\ref{eq:angularvelocity}).
In this case, Eq. (\ref{eq:kerrsen}) takes the form 

\begin{equation}
ds^{2}=-F\left(r\right)dt^{2}+\frac{1}{G\left(r\right)}dr^{2}+\rho^{2}d\theta^{2}+H^{2}\left(r\right)d\phi^{2}\label{eq:NewKerrSen}
\end{equation}
where

\begin{equation}
F\left(r\right)=\frac{\Delta\rho^{2}}{\left(r\left(r+r_{\alpha}\right)+a^{2}\right)^{2}-\Delta a^{2}\sin^{2}\theta}
\end{equation}

\begin{equation}
G\left(r\right)=\frac{\Delta}{\rho^{2}}
\end{equation}

\begin{equation}
H^{2}\left(r\right)=\frac{\sin^{2}\theta}{\rho^{2}}\left[\left(r\left(r+r_{\alpha}\right)+a^{2}\right)^{2}-\Delta a^{2}\sin^{2}\theta\right].
\end{equation}

\section{Area spectrum of a Sen black hole}

The Klein Gordon equation, 

\begin{equation}
g^{\mu\nu}\partial_{\mu}\partial_{\nu}\Phi-\frac{m^{2}}{\hbar^{2}}\Phi=0,
\end{equation}

gives the scalar field wave function $\Phi$ by using the metric \ref{eq:kerrsen}.
However, we can also obtain the wave function with the Hamilton-Jacobi
equation 
\begin{equation}
g^{\mu\nu}\partial_{\mu}S\partial_{\nu}S+m^{2}=0,
\end{equation}
where the action $S$ and the wave function$\Phi$ are related by
\begin{equation}
\Phi=\exp[\frac{i}{\hbar}S(t,r,\theta,\phi)].
\end{equation}
Now, we will concentrate on using the Hamilton-Jacobi equation to
find the wave function. In the dragging coordinate frame, the action
$S$ can be decomposed as \cite{Zeng2012,Chen665}

\begin{equation}
S(t,r,\theta)=-(E-m\Omega)t+W(r)+\Theta(\theta),
\end{equation}
where $E$ is the energy of the emitted particle measured by the observer
at the infinity and $m$ denotes the angular quantum number about
$\phi$. Using Eqs.(\ref{eq:kerrsen}) and (\ref{eq:NewKerrSen}),
we find that near the horizon $\Theta$ vanishes and $W$ can be solved
as

\begin{equation}
W(r)=\frac{i\pi(E-m\Omega)}{\sqrt{F^{\prime}(r_{H})G^{\prime}(r_{H})}},
\end{equation}
where we only consider the outgoing wave and primes denote derivatives
with respect to $r$. Hence, the wave function $\Phi$ can be written
in the form 

\begin{equation}
\Phi=\exp[-\frac{i}{\hbar}(E-m\Omega)t]\psi(r_{H})
\end{equation}

where $\psi(r_{H})=\exp[-\frac{\pi(E-m\Omega)}{\sqrt{F^{\prime}(r_{H})G^{\prime}(r_{H})}}]$.
Note that $\Phi$ is a periodic function with the period 
\begin{equation}
T=\frac{2\pi}{\left(\frac{E-m\Omega}{\hbar}\right)}
\end{equation}

or taking into account the relation $E=\hbar\omega$,

\begin{equation}
T=\frac{2\pi}{\left(\omega-\frac{m\Omega}{\hbar}\right)}.\label{eq:temperature2}
\end{equation}

It is well known that in Kruskal coordinates, the gravity system is
a periodic system with respect to the Euclidean time. Hence, we will
assume that particles moving in this background also own a period
that has a geometric origin in the Hawking thermal radiation \cite{Gibbons1978}.
Therefore, the relation between $T$ and $T_{H}$ is 
\begin{equation}
T=\frac{2\pi}{\kappa_{H}}=\frac{\hbar}{T_{H}}.\label{eq:temperaturerelation}
\end{equation}
Based on Eq. (\ref{eq:area}), the change of horizon area of a Kerr
black hole can be written as

\begin{equation}
\Delta A=8\pi\left[\frac{\left(\sqrt{\left(2M^{2}-Q^{2}\right)^{2}-4J^{2}}+2M^{2}-Q^{2}\right)\left(2MdM-QdQ\right)-2JdJ}{\sqrt{\left(2M^{2}-Q^{2}\right)^{2}-4J^{2}}}\right]
\end{equation}

or using Eq. (\ref{eq:HawkingTemperature}),

\begin{equation}
\Delta A=8\pi\left[\frac{\hbar\left(2MdM-QdQ\right)}{4\pi MT_{H}}-\frac{2JdJ}{\sqrt{\left(2M^{2}-Q^{2}\right)^{2}-4J^{2}}}\right].\label{eq:deltaA}
\end{equation}

Since $M$ is the total mass of the black hole and using Eqs. (\ref{eq:temperature2})
and (\ref{eq:temperaturerelation}), we can write 
\begin{equation}
dM-VdQ=\hbar\omega=m\Omega+2\pi T_{H}.
\end{equation}

Substituting this relation into Eq.(\ref{eq:deltaA}) gives

\begin{equation}
\Delta A=8\pi\left[\hbar+\frac{\hbar m\Omega}{2\pi T_{H}}+\frac{\hbar VdQ}{2\pi T_{H}}-\frac{\hbar QdQ}{4\pi MT_{H}}-\frac{2JdJ}{\sqrt{\left(2M^{2}-Q^{2}\right)^{2}-4J^{2}}}\right]
\end{equation}
and simplifying this expression using Eqs. (\ref{eq:angularvelocity})
and (\ref{eq:electricpotential}), we finally obtain 

\[
\Delta A=8\pi l_{p}^{2}.
\]
Note that the equally spaced area spectrum of a Sen black hole obtained
here is consistent with the result presented from the viewpoint of
quasinormal modes for the GMGHS black hole in \cite{Wei2010} and
for the Kerr black hole in \cite{Vagenas0811,Medved25,Zeng2012}.
However, the small angular momentum limit, which is necessary from
the perspective of quasinormal mode analysis, is not necessary to
obtain the general area gap $8\pi l_{p}^{2}$.

\section{Conclusions}

Although the quantum gravity theory has not been found, it is meaningful
to investigate the quantum correction to the area spectrum. We use
the new scheme proposed by Zeng et. al. \cite{Zeng2012} to quantize
the horizon area of a charged rotating black hole of heterotic string
theory. It was found that the period of the gravity system with respect
to the Euclidean time can determine the area spectrum of black holes.
This result confirms the speculation of Maggiore that the periodicity
of a black hole may be the origin of the area quantization. It is
important to note that this approach is more convenient and simple
to apply to the Sen metric since the quasinormal mode frequency could
lead to some confusion on whether the real part or imaginary part
is responsible for the area spectrum. 

\emph{Acknowledgements}

This work was supported by the Universidad Nacional de Colombia. Hermes
Project Code 13038.

\end{document}